\documentclass[%
 aip,
 amsmath,amssymb,
 reprint,%
]{revtex4-2}

\usepackage{graphicx}
\usepackage{dcolumn}
\usepackage{bm}
\usepackage{xcolor}

\usepackage[utf8]{inputenc}
\usepackage[T1]{fontenc}
\usepackage{mathptmx}
\usepackage{etoolbox}
\usepackage[colorlinks=true, allcolors=blue]{hyperref}
\makeatletter
\def\@email#1#2{%
 \endgroup
 \patchcmd{\titleblock@produce}
  {\frontmatter@RRAPformat}
  {\frontmatter@RRAPformat{\produce@RRAP{*#1\href{mailto:#2}{#2}}}\frontmatter@RRAPformat}
  {}{}
}%
\makeatother
\begin{document}

\preprint{AIP/123-QED}

\title[~]{Revisiting the epitaxial Si$_3$N$_4$ crystalline cap on AlGaN/GaN via evolutionary structure search}
\author{Xin Chen}
\affiliation{Thermal Science Research Center, Shandong Institute of Advanced Technology, Jinan 250100, Shandong Province, China}

\author{Xin Luo}
\affiliation{ 
Institute of Novel Semiconductors, Shandong University, Jinan 250100, China 
}%

\author{Duo Wang}
\affiliation{ 
Faculty of Applied Sciences, Macao Polytechnic University, Macao SAR, 999078, People’s Republic of China
}

\author{Xu Cheng}
\email{xchengab@sdu.edu.cn}
\affiliation{
Institute for Advanced Technology, Shandong University, Jinan 250061, People's Republic of China
}

\author{Peng Cui}
\email{pcui@sdu.edu.cn}
\affiliation{ 
Institute of Novel Semiconductors, Shandong University, Jinan 250100, China 
}%

\date{\today}

\begin{abstract}
In our recent experimental work (\textit{Appl. Phys. Lett. 125, 122109 (2024)}), we observed that crystalline Si$_3$N$_4$ cap layers a few nanometers thick can form \textit{in situ} on GaN surfaces. Compared with amorphous SiO$_2$ and Al$_2$O$_3$ caps, these crystalline caps yield cleaner GaN/Si$_3$N$_4$ interfaces with fewer defects and improved device metrics. These observations motivate two questions: why does Si$_3$N$_4$ away from the interface become amorphous as the cap thickens, and what is the actual crystal structure of the interfacial Si$_3$N$_4$? Prior work proposed a defect-wurtzite (DW) model constructed heuristically from $\beta$-Si$_3$N$_4$ and the AlGaN lattice constants, but it is significantly higher in energy than $\beta$-Si$_3$N$_4$ and disagrees with experiment in both interlayer spacings and electronic gap. Using a systematic structure-search approach under in-plane lattice constraints commensurate with AlGaN, we identify a lower-energy configuration, denoted Lam-Si$_3$N$_4$, with quasi-two-dimensional (laminar) stacking normal to the interface. Under AlGaN-matched metrics, Lam-Si$_3$N$_4$ is about 60 meV/atom more stable than DW-Si$_3$N$_4$ and reproduces the experimentally observed interlayer spacings more closely. The substantial lattice mismatch explains amorphization when the crystalline cap grows far from the interface. Upon full relaxation, both DW- and Lam-Si$_3$N$_4$ exhibit wide $\sim$4 eV band gaps. Under AlGaN constraints, the DW gap collapses to $\sim$1.88 eV whereas Lam-Si$_3$N$_4$ maintains a larger $\sim$2.70 eV gap (for reference, PBE gaps: GaN 1.73 eV, AlN 4.05 eV). The wider gap and improved structural match of Lam-Si$_3$N$_4$ rationalize the superior capping performance and provide guidance for optimizing AlGaN/GaN device encapsulation.

\end{abstract}

\maketitle

Gallium nitride (GaN)-based high electron mobility transistors (HEMTs) have shown significant promise for high-frequency and high-power electronic applications due to their outstanding properties, such as wide bandgap, high electron saturation velocity, and high breakdown electric field \cite{10128694,8962057,MOUNIKA2022207317,AJAYAN2022106982}. A critical component to optimize the performance and reliability of GaN-based HEMTs is the integration of a thin cap layer on the AlGaN barrier, typically made from silicon nitride (Si$_3$N$_4$) \cite{Liu_2011}. Such Si$_3$N$_4$ cap layers are important for reducing surface trap states, enhancing polarization effects, and improving device stability  \cite{10.1063/1.122312}.

Historically, Si$_3$N$_4$ films deposited by \textit{ex situ} methods, like low-pressure chemical vapor deposition (LPCVD)\cite{7370783,10.1186/1556-276X-9-474}, have dominated due to ease of processing. However, these methods introduce high densities of interface states and defects, negatively impacting device performance. Recent advancements in metal-organic chemical vapor deposition (MOCVD) techniques have enabled the growth of high-quality crystalline Si$_3$N$_4$ directly onto AlGaN, providing a substantial reduction in interface states and improved electrical characteristics \cite{10.1063/5.0224144,PENG2025163416}.

An initial study proposed a defect-wurtzite structure (DW-Si$_3$N$_4$) for these epitaxially grown films, supported by first-principles calculations and high-resolution transmission electron microscopy (HRTEM) observations \cite{Takizawa2008}. Nevertheless, although plausible, the DW-Si$_3$N$_4$ structure exhibits higher formation energies and discrepancies in matching experimental lattice constants, highlighting the need for a more systematic investigation.

In our recent experimental work \cite{10.1063/5.0224144}, we synthesized a notably thicker ($\sim$2~nm) crystalline Si$_3$N$_4$ layer on the AlGaN barrier using an optimized MOCVD growth method. This advancement enhanced the performance of GaN-based HEMTs, exhibiting higher electron mobility, improved current density, and substantially higher breakdown voltages. Motivated by these promising results, we undertook a global structure search using evolutionary algorithms to identify a more energetically favorable configuration of Si$_3$N$_4$.

Reliable structural predictions are pivotal for guiding the synthesis of new materials and understanding their functional behavior. While experimental measurements, such as lattice constants, offer valuable initial insights, they often fall short in providing the detailed structural information needed to accurately predict material behavior at the atomic scale. In recent years, several algorithms have been proposed to address this fundamental challenge, including simulated annealing, genetic algorithms, data mining, and metadynamics. Among these methods, evolutionary algorithms, where a population of candidate structures evolves through iterations of random variation and selection, have proven to be particularly effective \cite{doi:10.1063/1.2210932,GLASS2006713}. The application of evolutionary algorithms has facilitated the prediction of numerous new materials characterized by excellent stability and intriguing properties\cite{Zhang1502,Oganov2019,PhysRevLett.112.085502,PhysRevB.103.075429,PhysRevB.87.195317,PhysRevLett.110.136403,Zhang2017,PhysRevB.111.155425}.

\begin{figure}[htbp]
\includegraphics[scale=1]{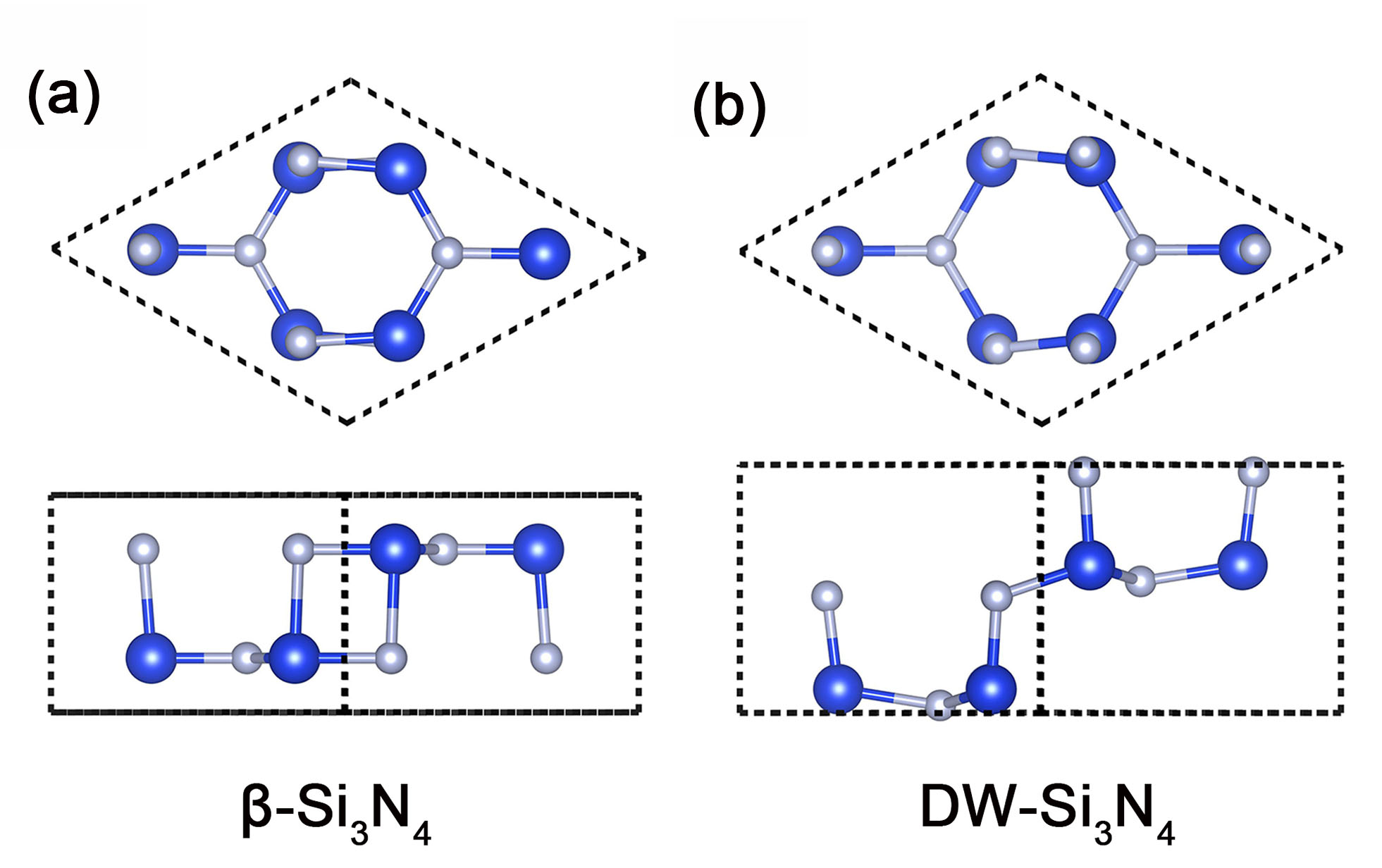} 
\caption{(a) Top and (b) side views of $\beta$-Si$_3$N$_4$ and DW-Si$_3$N$_4$.}
\label{figure1}
\end{figure}

In this work, motivated by our observation that few‐nanometer \textit{in situ} crystalline Si$_3$N$_4$ caps on AlGaN/GaN yield cleaner interfaces and better device metrics than amorphous SiO$_2$ or Al$_2$O$_3$, we revisit the atomic structure of the cap layer and its thickness dependence. Two questions guide our study: (i) why Si$_3$N$_4$ away from the interface tends to become amorphous as the cap gets thicker, and (ii) what crystalline configuration best represents the interfacial Si$_3$N$_4$. To answer these questions, we carry out an extensive evolutionary structure search under in‑plane lattice constants matched to AlGaN, exploring 691 candidate configurations of Si$_3$N$_4$ on AlGaN(0001). We identify a quasi‑two‑dimensional, laminar phase, denoted Lam‑Si$_3$N$_4$, which is lower in energy by $\sim 60$~meV/atom than the previously proposed DW model under the same epitaxial constraint and reproduces the experimentally observed interlayer spacings more closely. The large lattice mismatch provides a natural explanation for the thickness‑dependent crystallinity: ordering is favored near the interface, while regions farther from the interface more readily lose registry and become amorphous. We further compare the electronic structures, mechanical response, and lattice dynamics of the DW and Lam phases. When fully relaxed, both exhibit wide band gaps near 4~eV, consistent with insulating behavior. Under the AlGaN‑matched constraint, however, the DW gap shrinks to about 1.9~eV, whereas Lam‑Si$_3$N$_4$ maintains a larger gap of about 2.6~eV. For reference, PBE gaps for GaN and AlN are 1.73~eV and 4.05~eV, respectively. The ability of Lam‑Si$_3$N$_4$ to preserve a wider gap under epitaxial strain, together with its lower energy and better structural match, rationalizes the superior dielectric performance of crystalline Si$_3$N$_4$ caps and points to actionable routes for optimizing GaN HEMT capping layers.

\begin{figure}[htbp]
\includegraphics[scale=1]{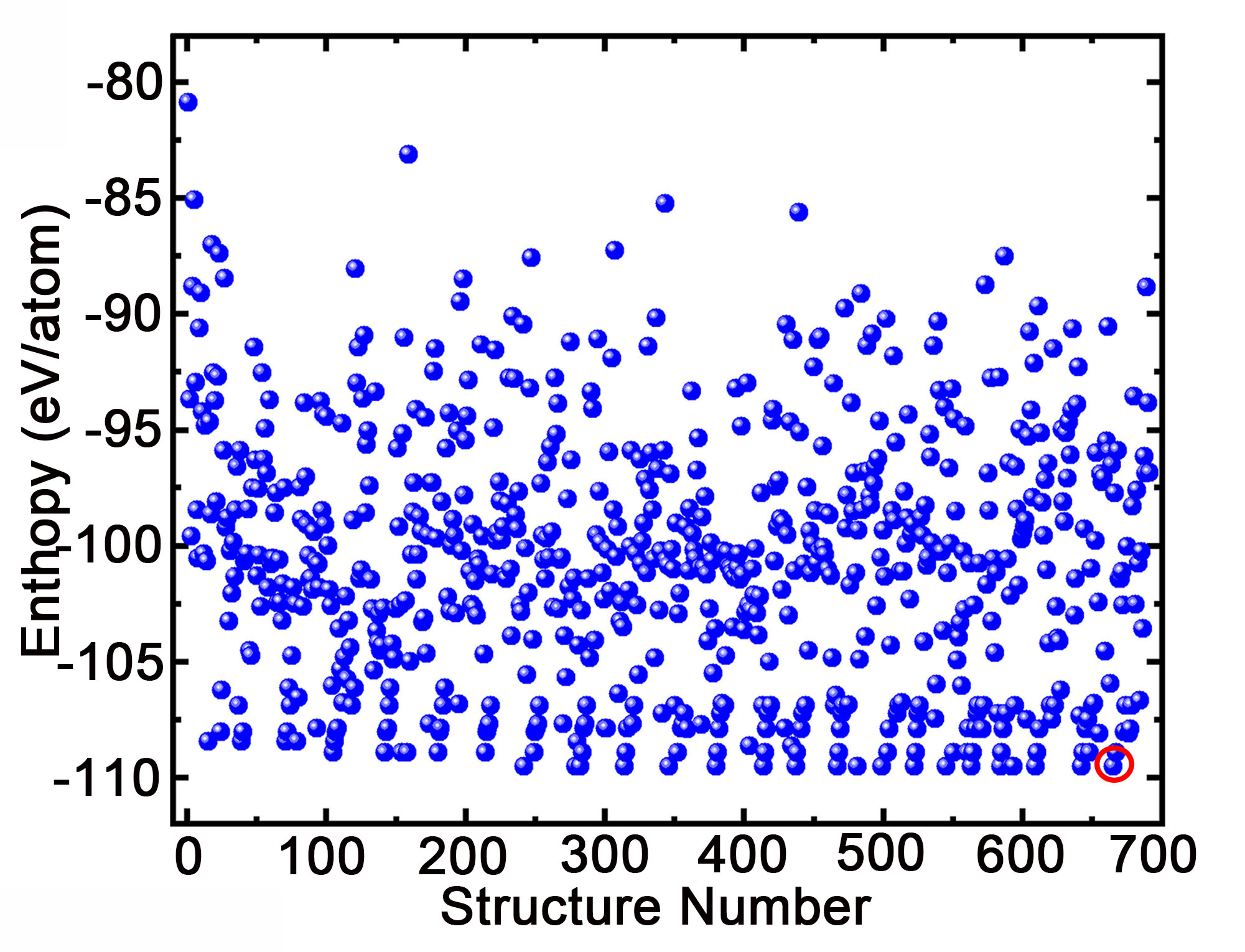} 
\caption{The enthalpies of the allotropes found in the evolutionary structure search. In the red circle is Lam-Si$_3$N$_4$. }
\label{figure2}
\end{figure}

We performed a global structural search using the evolutionary algorithm-based code \textsc{USPEX}~\cite{doi:10.1063/1.2210932,GLASS2006713,PhysRevB.87.195317,doi:10.1021/ar1001318,LYAKHOV20131172} to explore low-energy configurations of Si$_3$N$_4$. The simulation cell was constrained with lattice parameters of (6.4, 6.4, 4.8)~\AA~and lattice angles of (90$^\circ$, 90$^\circ$, 120$^\circ$), corresponding to a hexagonal symmetry. Each unit cell was set to contain 6 Si atoms and 8 N atoms. A total of 691 structures were generated and evaluated during the search. The structural search was considered converged when the lowest-enthalpy structure remained unchanged for 10 generations. First-principles calculations were performed using the Vienna \textit{Ab initio} Simulation Package (\textsc{VASP})~\cite{002230939500355X,PhysRevB.54.11169}, with the generalized gradient approximation (GGA) in the form of Perdew, Burke, and Ernzerhof (PBE)~\cite{PhysRevLett.77.3865} for the exchange-correlation functional. The plane-wave energy cutoff was set to 520~eV, and convergence criteria were set to $10^{-6}$~eV for energy and 0.001~eV/\AA for atomic forces. Brillouin zone integration was performed using a Monkhorst-Pack grid denser than $2\pi \times 0.03$~\AA$^{-1}$. To assess dynamical stability, phonon spectra were computed using the finite displacement method via \textsc{PHONOPY}~\cite{phonopy}. Post-processing, including geometry structure, band structure and phonon spectra analysis, was conducted using \textsc{VASPKIT}~\cite{wang2019vaspkit} and visualized with \textsc{VESTA}~\cite{Mommako5060}.

\begin{figure}[htbp]
\includegraphics[scale=1]{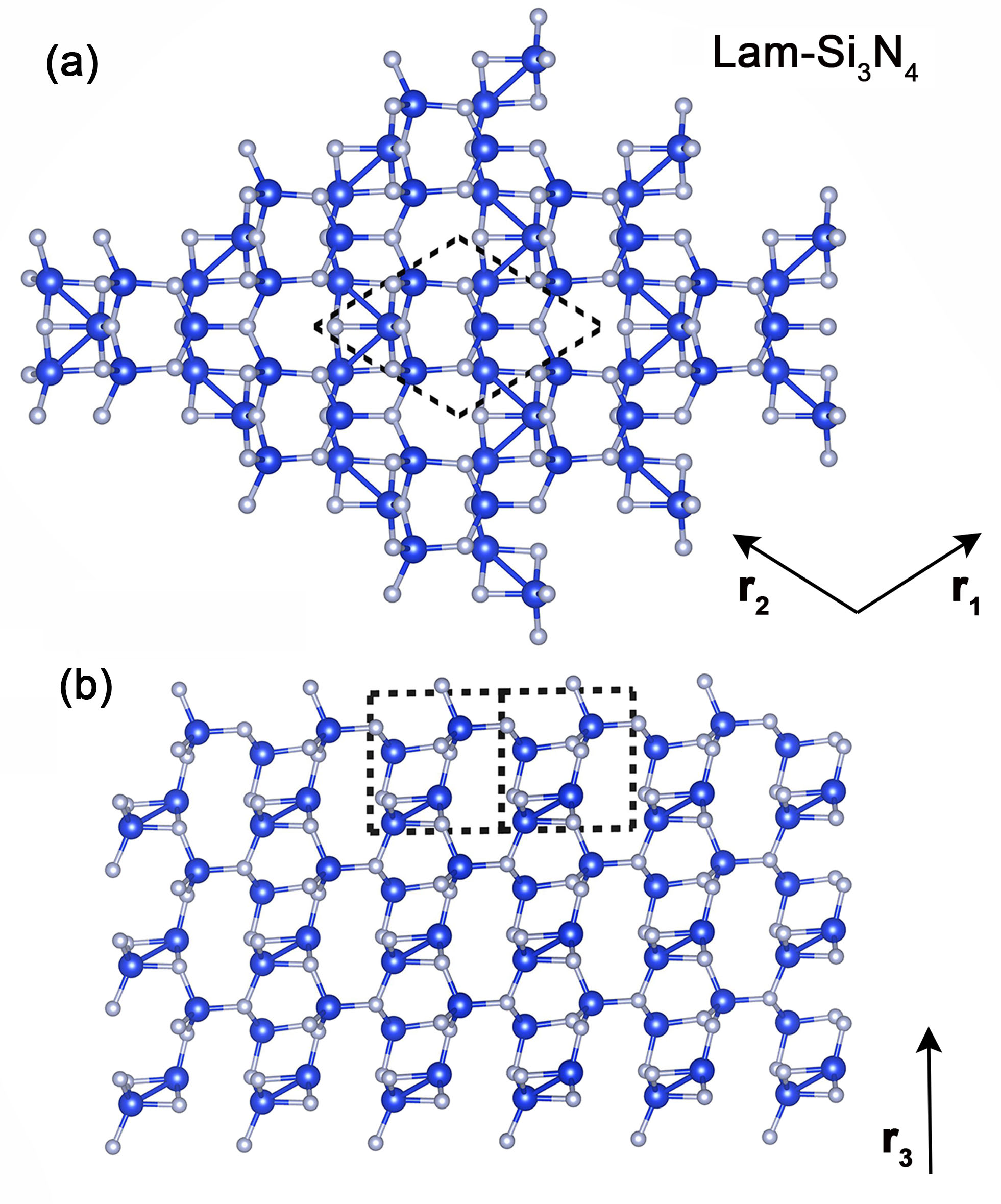} 
\caption{(a) Top and (b) side views of Lam-Si$_3$N$_4$. White and blue spheres represent N and Si atoms, respectively.}
\label{figure3}
\end{figure}

Ref. \cite{10.1063/5.0224144} Fig. 1 shows a high-resolution cross-sectional TEM image of the Si$_3$N$_4$/AlGaN/AlN/GaN heterostructure. Clear interfaces are observed between the different epitaxial layers. The measured thicknesses of the Si$_3$N$_4$, AlGaN, and AlN layers are 1.82~nm, 18.04~nm, and 0.96~nm, respectively. The TEM specimen was prepared using a focused ion beam (FIB) technique before device fabrication. In the enlarged view of the Si$_3$N$_4$/AlGaN interface, a well-ordered atomic arrangement is observed, with the vertical lattice spacing measured to be 0.4780~nm for the Si$_3$N$_4$ layer and 0.5068~nm for the AlGaN layer along the vertical direction. The coherent lattice match and absence of visible dislocations indicate that the in situ Si$_3$N$_4$ layer possesses a crystalline structure and grows epitaxially on AlGaN.

Although $\beta$-Si$_3$N$_4$, as shown in Fig.~\ref{figure1}(a), is the most thermodynamically stable phase in bulk form, it has been found to be inconsistent with experimental observations in epitaxial Si$_3$N$_4$ layers grown on AlGaN. The lattice constant along the $c$-axis in the $\beta$-Si$_3$N$_4$ structure is significantly smaller than that measured from high-resolution TEM. In contrast, the DW structure of Si$_3$N$_4$, as shown in Figure~\ref{figure1}(b), while sharing a similar bonding network with the $\beta$ phase, exhibits an expanded out-of-plane lattice constant, more closely matching the experimentally observed interplanar spacing of 0.478~nm. This better agreement in lattice parameters made DW-Si$_3$N$_4$ a reasonable candidate structure in earlier theoretical models. However, the DW phase was proposed based on empirical reasoning in the absence of global structure search techniques at the time. As such, whether it represents the most reasonable atomic configuration remained an open question. In particular, although the out-of-plane lattice constant $c$ in the DW structure matches experimental measurements better than that of the $\beta$ phase, we found that when fixing the in-plane lattice constants $a$ and $b$ to experimental values, the optimized $c$ value of DW-Si$_3$N$_4$ still shows a noticeable deviation from the measured result. This discrepancy highlighted the need for a systematic and unbiased structural search to identify more accurate low-energy configurations of Si$_3$N$_4$ under epitaxial constraints.

To determine the most stable atomic configuration of Si$_3$N$_4$, we conducted a comprehensive structure search using an evolutionary algorithm. Fig.~\ref{figure2} presents the energy distribution of 691 candidate structures, with each structure relaxed under the same lattice constant constraints. Among them, a novel phase—denoted as Lam-Si$_3$N$_4$—is identified as the lowest energy structure. 

The representative atomic arrangements of Lam-Si$_3$N$_4$ are shown in Fig.~\ref{figure3}. While $\beta$- and DW-Si$_3$N$_4$ exhibit conventional three-dimensional frameworks, the Lam-Si$_3$N$_4$ structure is characterized by a quasi-two-dimensional (laminar) configuration with distinct in-plane ($\mathbf{r}_1 +\mathbf{r}_2$, $\mathbf{r}_3$) and out-of-plane ($\mathbf{r}_1 -\mathbf{r}_2$) anisotropy. The Lam-Si$_3$N$_4$ phase crystallizes in a monoclinic system (space group $Cm$). The Si and N atoms form a layered network, suggesting strong in-plane bonding and potentially anisotropic physical properties. The crystalline nature of the in situ grown Si$_3$N$_4$ layer observed in Ref. \cite{10.1063/5.0224144} Fig. 1, along with the energetically favorable formation of Lam-Si$_3$N$_4$, provides compelling evidence that this structure may represent the true atomic configuration of Si$_3$N$_4$ grown on AlGaN.

\begin{figure}[htbp]
\includegraphics[scale=1.03]{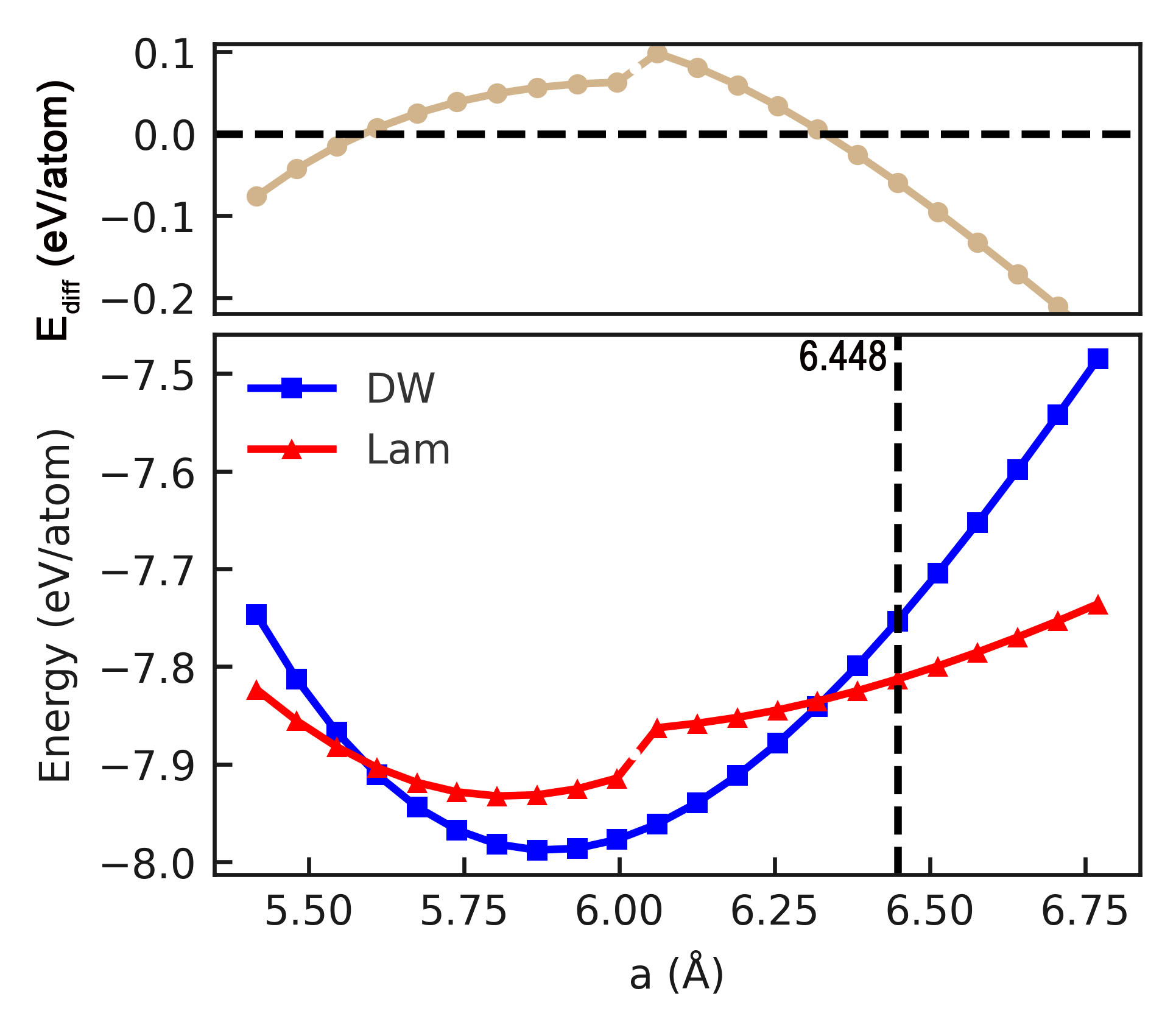} 
\caption{Total energies of fully-relaxed DW-Si$_3$N$_4$ and Lam-Si$_3$N$_4$ versus the in-plane lattice constant $a$. The bottom panel shows total energies, while the top panel shows the energy difference $E_{\mathrm{Lam}} - E_{\mathrm{DW}}$. Under the fully-relaxed AlGaN in-plane lattice constant of $\sim$6.448~\AA{} (dashed line), Lam-Si$_3$N$_4$ is favored by $\sim$60~meV/atom compared to DW-Si$_3$N$_4$.}
\label{figure4}
\end{figure}

To further evaluate the stability of the DW and Lam phases under epitaxial constraints, we calculated their total energies as a function of the in-plane lattice constant $a$, as shown in Figure~\ref{figure4}. The bottom panel presents the absolute energies of both structures, while the top panel highlights the energy difference between them ($E_{\mathrm{diff}} = E_{\mathrm{Lam}} - E_{\mathrm{DW}}$). It is evident that the DW phase has a lower total energy than the Lam structure at their respective equilibrium lattice constants, with a minimum near $a \approx 5.80$~\AA. However, as $a$ increases beyond this value, the energy of the DW structure rises steeply, while that of the Lam phase increases more moderately.

Remarkably, at the fully relaxed AlGaN in-plane lattice constant of 6.448~\AA{}, which is indicated by the dashed vertical line in the Figure, the Lam phase becomes significantly more favorable, exhibiting an energy advantage of approximately 60~meV/atom over the DW structure. This energy difference is substantial, indicating that under experimental epitaxial conditions, the Lam-Si$_3$N$_4$ structure is thermodynamically preferred. Notably, PBE calculations are known to yield slightly overestimated bond lengths, and our computed lattice constant of 6.448~\AA{} agrees well with the experimental value of $\sim$6.4~\AA. These results underscore the limitation of the DW model in strained heteroepitaxial environments and support the assignment of the Lam phase as a more favorable candidate for \textit{in situ} grown crystalline Si$_3$N$_4$.

In addition, we observe a noticeable discontinuity in the total energy of the Lam structure around $a \approx 6.0$~\AA, as shown in Fig.~\ref{figure4}. Specifically, the energy does not vary smoothly with increasing lattice constant, but instead exhibits a distinct kink, indicating a first-order structural phase transition. This behavior suggests that the Lam structure undergoes a sudden reconstruction in its atomic configuration to accommodate the imposed in-plane strain. Such a transition implies the existence of two locally stable configurations within the Lam phase that differ in symmetry or bonding topology, and are separated by an energy barrier. The emergence of this strain-induced phase transition highlights the structural flexibility of Lam-Si$_3$N$_4$ under epitaxial constraints, in contrast to the relatively rigid framework of the DW structure. From a thermodynamic perspective, this adaptability enhances the likelihood that Lam-Si$_3$N$_4$ may form preferentially during in situ growth on lattice-mismatched substrates, further supporting its assignment as the more realistic candidate for the experimentally observed crystalline Si$_3$N$_4$ phase.

\begin{figure}[htbp]
\includegraphics[scale=1.02]{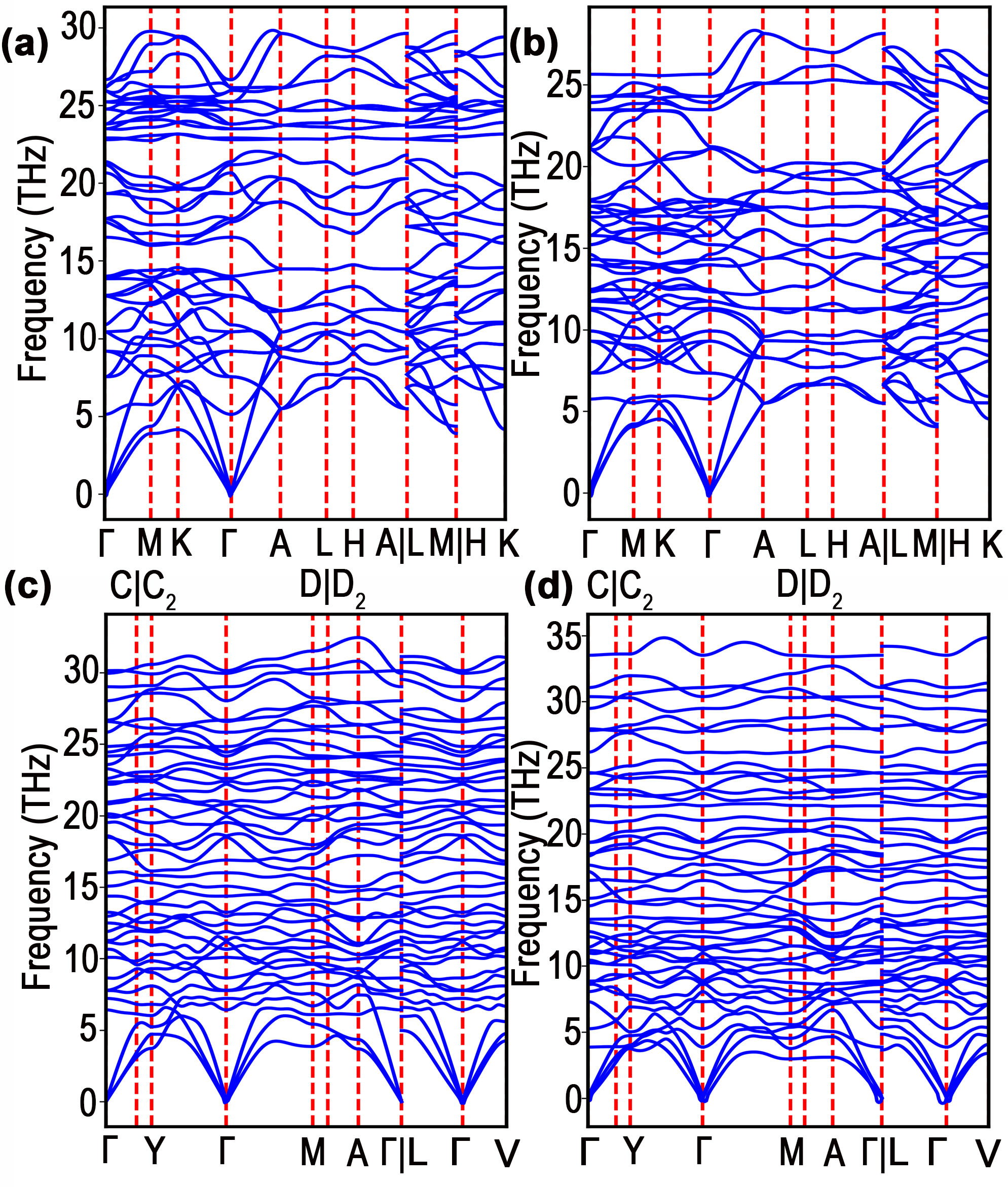}
\caption{Phonon spectra of (a) fully-relaxed DW-Si$_3$N$_4$, (b) DW-Si$_3$N$_4$ under lattice constant constraint, (c) fully-relaxed Lam-Si$_3$N$_4$, and (d) Lam-Si$_3$N$_4$ under lattice constant constraint. No imaginary frequencies are observed, indicating dynamic stability in all cases.}
\label{figure5}
\end{figure}

To assess the dynamical stability of both the DW-Si$_3$N$_4$ and Lam-Si$_3$N$_4$ structures, we calculated their phonon dispersion spectra using the finite displacement method as implemented in the \textsc{PHONOPY} code. A $2 \times 2 \times 2$ supercell and corresponding $\Gamma$-centered $k$-mesh were employed to ensure convergence. The resulting phonon spectra are shown in Figure~\ref{figure5}, where panels (a) and (b) correspond to the fully-relaxed DW structure and DW Si$_3$N$_4$ with a and b lattice constant fixed, respectively, and panels (c) and (d) correspond to the Lam structure under the same conditions. As shown in all four cases, the phonon branches are entirely real and no imaginary frequencies are observed throughout the Brillouin zone. This confirms that both DW-Si$_3$N$_4$ and Lam-Si$_3$N$_4$ structures are dynamically stable, whether fully relaxed or constrained to match the experimental in-plane lattice constants. In particular, the Lam structure retains its dynamical stability under the experimentally relevant epitaxial constraint, which complements its thermodynamic favorability discussed earlier. These results together establish Lam-Si$_3$N$_4$ as a mechanically and dynamically robust candidate phase for epitaxial growth on AlGaN surfaces.

Tables~\ref{tab:Cij-Lam} and \ref{tab:Cij-DW} list the complete $6\times6$ stiffness (elastic) matrices $C_{ij}$ for the two Si$_3$N$_4$ structures considered: Lam-Si$_3$N$_4$ and the hexagonal DW-Si$_3$N$_4$ (space group $P6_3mc$) \cite{Gaillac_2016}. For the Lam-Si$_3$N$_4$ structure, there are 13 independent elastic constants, while in the DW-Si$_3$N$_4$ system, there are 5 independent constants. These full matrices highlight the numerical values of each elastic-constant component, reflecting the different symmetry constraints of each crystal system.

\begin{table}[htbp]
\centering
\caption{\label{tab:Cij-Lam}
Complete stiffness matrix $C_{ij}$ (in GPa) for Lam-Si$_3$N$_4$.}
\begin{tabular}{c|cccccc}
\hline\hline
 & $j=1$ & $j=2$ & $j=3$ & $j=4$ & $j=5$ & $j=6$ \\
\hline
$i=1$ & 251.947 & 47.017  & 71.098  & 0.000   & 7.598   & 0.000   \\
$i=2$ & 47.017  & 348.053 & 101.089 & 0.000   & 4.407   & 0.000   \\
$i=3$ & 71.098  & 101.089 & 351.163 & 0.000   & 47.423  & 0.000   \\
$i=4$ & 0.000   & 0.000   & 0.000   & 96.011  & 0.000   & 16.625  \\
$i=5$ & 7.598   & 4.407   & 47.423  & 0.000   & 97.219  & 0.000   \\
$i=6$ & 0.000   & 0.000   & 0.000   & 16.625  & 0.000   & 87.656  \\
\hline\hline
\end{tabular}
\end{table}

\begin{table}[htbp]
\centering
\caption{\label{tab:Cij-DW}
Complete stiffness matrix $C_{ij}$ (in GPa) for DW-Si$_3$N$_4$.}
\begin{tabular}{c|cccccc}
\hline\hline
 & $j=1$ & $j=2$ & $j=3$ & $j=4$ & $j=5$ & $j=6$ \\
\hline
$i=1$ & 392.710 & 126.113 & 60.326  & 0.000   & 0.000   & 0.000   \\
$i=2$ & 126.113 & 392.710 & 60.326  & 0.000   & 0.000   & 0.000   \\
$i=3$ & 60.326  & 60.326  & 416.437 & 0.000   & 0.000   & 0.000   \\
$i=4$ & 0.000   & 0.000   & 0.000   & 82.368  & 0.000   & 0.000   \\
$i=5$ & 0.000   & 0.000   & 0.000   & 0.000   & 82.368  & 0.000   \\
$i=6$ & 0.000   & 0.000   & 0.000   & 0.000   & 0.000   & 133.299 \\
\hline\hline
\end{tabular}
\end{table}

A direct comparison of the diagonal elements reveals that the hexagonal phase (DW-Si$_3$N$_4$) generally has higher values for $C_{11}$ and $C_{33}$ (392.71\,GPa and 416.44\,GPa, respectively) than the corresponding diagonal terms in the monoclinic Lam-Si$_3$N$_4$ (where $C_{11} = 251.95$\,GPa, $C_{33}=351.16$\,GPa). Notably, $C_{66}$ in the hexagonal phase (133.30\,GPa) is also significantly larger than its counterpart in the monoclinic system (87.66\,GPa). By contrast, some of the off-diagonal terms in Lam-Si$_3$N$_4$, such as $C_{15}=7.598$\,GPa and $C_{35}=47.423$\,GPa, have no direct analog in the hexagonal structure due to symmetry. Another major difference is in $C_{12}$ (47.02\,GPa in Lam-Si$_3$N$_4$ vs.\ 126.11\,GPa in DW-Si$_3$N$_4$), suggesting distinct in-plane bonding characteristics.

\begin{figure*}[hbtp]
\includegraphics[scale=1.02]{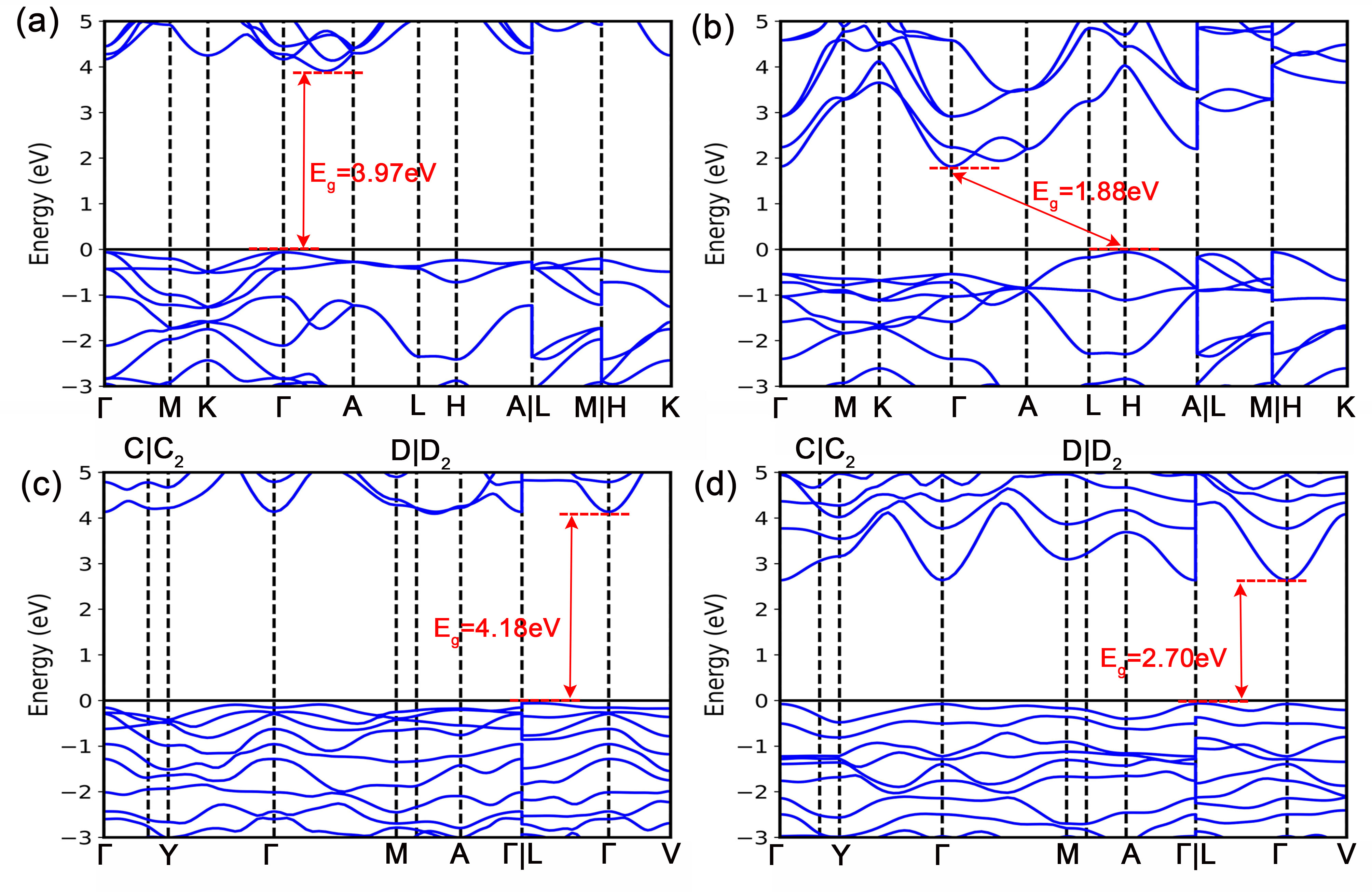}
\caption{Electronic band structures of (a) fully-relaxed DW-Si$_3$N$_4$, (b) DW-Si$_3$N$_4$ under lattice constant constraint, (c) fully-relaxed Lam-Si$_3$N$_4$, and (d) Lam-Si$_3$N$_4$ under lattice constant constraint. The Fermi level is set to zero. Note the significant gap reduction in DW-Si$_3$N$_4$ when under constraint, whereas Lam-Si$_3$N$_4$ maintains a relatively larger band gap.}
\label{figure6}
\end{figure*}

These variations in the stiffness matrix lead to differences in the computed macroscopic elastic moduli. For example, the Hill-averaged bulk modulus of Lam-Si$_3$N$_4$ is about 149.78\,GPa, whereas that of DW-Si$_3$N$_4$ is $\sim$188.20\,GPa, indicating stronger resistance to volumetric deformation in the hexagonal phase. The Young's modulus is likewise higher in DW-Si$_3$N$_4$ (up to $\sim$291.92\,GPa) compared to Lam-Si$_3$N$_4$ ($\sim$248.77\,GPa). Both materials exhibit Pugh's ratios ($B/G$) below 1.75, suggesting brittle mechanical behavior common to many ceramic materials. 

Furthermore, the calculated longitudinal and transverse sound velocities for DW-Si$_3$N$_4$ (10\,316\,m/s and 6\,023\,m/s, respectively) slightly exceed those of Lam-Si$_3$N$_4$ (9\,577\,m/s and 5\,717\,m/s), consistent with the higher stiffness observed in the hexagonal phase. Correspondingly, the Debye temperature of DW-Si$_3$N$_4$ is estimated to be about 915\,K, compared to about 855\,K for Lam-Si$_3$N$_4$, implying potentially higher thermal conductivity in the hexagonal material.

We further verified the standard elastic-stability criteria for both phases to check the mechanical stability. Monoclinic Lam-Si$_3$N$_4$ satisfies requirements such as $C_{33}C_{55} - C_{35}^2 > 0$ and $C_{44}C_{66} - C_{46}^2 > 0$, while DW-Si$_3$N$_4$ meets the hexagonal criteria $C_{11} > |C_{12}|$, $2C_{13}^2 < C_{33}(C_{11}+C_{12})$, and $C_{44} > 0$. All relevant inequalities are fulfilled, and the eigenvalues of the stiffness matrices are positive in each case, confirming the mechanical stability of both structures.

In addition to the overall stiffness differences, the two Si$_3$N$_4$ phases also exhibit contrasting degrees of anisotropy. For single crystals, one measure of anisotropy is the range between the minimum and maximum directional elastic moduli, such as the Young’s modulus $E(\hat{n})$ in different crystallographic directions $\hat{n}$. In Lam-Si$_3$N$_4$, the computed minimum and maximum values of $E$ are approximately 190\,GPa and 350\,GPa, respectively, yielding an anisotropy factor of $\sim$1.84. By contrast, DW-Si$_3$N$_4$ shows a somewhat broader $E$ range from 236\,GPa to 402\,GPa, corresponding to an anisotropy factor of $\sim$1.70. Similar trends are observed for the shear modulus $G$ and Poisson’s ratio $\nu$, indicating that both phases are moderately anisotropic, although the specific patterns of directional dependence differ due to their distinct symmetries.

Another widely used metric is the universal elastic anisotropy index $A^U$. For Lam-Si$_3$N$_4$, $A^U$ is computed to be about 0.40, whereas DW-Si$_3$N$_4$ has a slightly higher value of around 0.51. Both are significantly above 0, confirming that these materials deviate from perfect isotropy. However, the lower symmetry of the monoclinic phase often implies more complex anisotropic behavior in certain off-diagonal elements. In contrast, the hexagonal symmetry of DW-Si$_3$N$_4$ concentrates anisotropy into fewer independent constants but still leads to strong directional variation in properties like the shear modulus ($G_{4}=G_{5}$ versus $G_{6}$). These findings illustrate that, despite the overall higher rigidity of the DW-Si$_3$N$_4$ phase, both structures show non-negligible elastic anisotropy arising from their crystallographic arrangements and bonding networks.


To investigate the electronic properties of DW-Si$_3$N$_4$ and Lam-Si$_3$N$_4$, we calculated their band structures using the PBE functional within the generalized gradient approximation. The results are shown in Fig.~\ref{figure6}, where panels (a) and (b) correspond to the DW structure in its fully relaxed and strained states, respectively, and panels (c) and (d) correspond to the Lam structure under the same two conditions. The Fermi level is aligned to zero in each panel for reference.

For the fully relaxed DW structure, as shown in Fig.~\ref{figure6}(a), we find an indirect band gap of 3.97~eV, with the valence band maximum (VBM) located at the $\Gamma$ point and the conduction band minimum (CBM) positioned near the $M$ point. Upon applying in-plane strain [Fig.~\ref{figure6}(b)] by fixing the $a$ and $b$ lattice constants to match experimental values, the band structure of DW-Si$_3$N$_4$ undergoes significant deformation. The band gap narrows to 1.88~eV and remains indirect. In addition, the curvature of the conduction bands increases, suggesting a reduced effective mass for electrons. Such changes imply potential improvement in carrier mobility. However, the corresponding reduction in band gap may increase leakage current if DW-Si$_3$N$_4$ were used as a dielectric cap in GaN-based devices, possibly compromising its insulating performance.

In contrast, the Lam-Si$_3$N$_4$ structure maintains a more stable electronic profile. In the fully relaxed configuration [Fig.~\ref{figure6}(c)], it shows an indirect band gap of 4.18~eV, with the VBM at $\Gamma$ and the CBM located near the $K$ point. When subjected to the same strain condition [Fig.~\ref{figure6}(d)], the Lam structure still preserves an indirect band gap of 2.7~eV with only minor shifts in band edge positions. The band dispersion remains smooth, and the overall electronic structure is minimally perturbed. This insensitivity to strain implies that Lam-Si$_3$N$_4$ can retain stable dielectric behavior even under lattice mismatch with AlGaN, consistent with experimental observations of robust electrical properties in devices employing in situ crystalline Si$_3$N$_4$ cap layers.

Moreover, the preservation of a wide band gap in Lam-Si$_3$N$_4$ under strain is consistent with the experimentally observed low interface leakage currents and strong gate modulation in GaN HEMTs capped with crystalline Si$_3$N$_4$. By contrast, the strain-induced electronic fluctuations predicted for DW-Si$_3$N$_4$ could compromise electrical isolation and enhance defect sensitivity, outcomes that are clearly at odds with experimental evidence. When considered together with thermodynamic and phonon stability analyses, these results reinforce Lam-Si$_3$N$_4$ as a more reliable structural model for high-performance epitaxial integration, in agreement with experimental demonstrations of its effectiveness as a capping layer.

In summary, we performed an evolutionary structure search under the in-plane lattice constraint of AlGaN and identified a quasi‑two‑dimensional laminar phase of Si$_3$N$_4$ (Lam‑Si$_3$N$_4$) as a more suitable interfacial structural model. At the fully-relaxed AlGaN in‑plane constant, Lam‑Si$_3$N$_4$ is lower in energy than the defect‑wurtzite model by $\sim 60$~meV/atom and better reproduces the measured interlayer spacing of the crystalline cap. Both Lam and DW phases are dynamically stable under the epitaxial constraint. The electronic response to constraint differentiates the two structures: the band gap of DW‑Si$_3$N$_4$ collapses to about $1.88$~eV, whereas Lam‑Si$_3$N$_4$ maintains a wider $\sim 2.7$~eV gap. Upon full relaxation, both recover wide gaps are near $4$~eV (3.97 eV and 4.18 eV respectively). These results rationalize the superior dielectric behavior observed for \textit{in situ} crystalline Si$_3$N$_4$ caps on AlGaN and explain the thickness‑dependent crystallinity expected under large lattice mismatch, with ordering favored near the interface and amorphization more likely away from it. The wider gap and improved structural match of the Lam-Si$_3$N$_4$ model rationalize the experimentally observed capping behavior and provide guidance for optimizing epitaxial strategies in AlGaN/GaN devices.

\begin{acknowledgments}
We would like to thank the Shandong Institute of Advanced Technology for providing computational resources. Xin Chen acknowledges financial support from the National Natural Science Foundation of Shandong Province(Grant No. ZR2024QA040). Duo Wang acknowledges financial support from the Science and Technology Development Fund from Macau SAR (Grant Nos. 0062/2023/ITP2 and 0016/2025/RIA1) and the Macao Polytechnic University (Grant No. RP/FCA-03/2023). Peng Cui acknowledges financial support from the National Natural Science Foundation of China (62304123), Excellent Youth Science Foundation of Shandong Province (2023HWYQ-50), and Natural Science Foundation of Shandong Province (ZR2023QF061).
\end{acknowledgments}

\section*{AUTHOR DECLARATIONS}
\subsection*{Conflict of Interest}
The authors have no conflicts to disclose.

\section*{Data Availability Statement}

The data that support the findings of this study are available
from the corresponding author upon reasonable request.


\nocite{*}
\bibliography{ref}

\end{document}